\documentclass[11pt, a4paper, fleqn]{article}

\usepackage{fullpage}
\usepackage[tbtags]{amsmath}
\usepackage{amssymb}
\usepackage{mathrsfs}

\newcommand{\op}[1]{\boldsymbol{#1}}
\newcommand{\opgr}[1]{\boldsymbol{#1}}
\newcommand{\nc}[1]{\widehat{#1}}

\begin{document}

\title{\textbf{Noncommutative gauge theories\\ and Lorentz symmetry}}

\author{{\normalsize \textbf{Rabin Banerjee, Biswajit Chakraborty and Kuldeep Kumar}}\\[1.8ex]
{\normalsize \textit{S. N. Bose National Centre for Basic Sciences,}}\\
{\normalsize \textit{JD Block, Sector 3, Salt Lake, Kolkata 700098, India}}\\[0.7ex]
{\normalsize {E-mail:} \texttt{rabin@bose.res.in}, \texttt{biswajit@bose.res.in}, \texttt{kuldeep@bose.res.in}}
}

\date{}

\maketitle

\begin{abstract}
We explicitly derive, following a Noether-like approach, the criteria for preserving Poincar\'e invariance in noncommutative gauge theories. Using these criteria we discuss the various spacetime symmetries in such theories. It is shown that, interpreted appropriately, Poincar\'e invariance holds. The analysis is performed in both the commutative as well as noncommutative descriptions and a compatibility between the two is also established.
\end{abstract}

\centerline{{\small PACS: 11.30.Cp, 11.10.Nx, 11.15.-q}}

\bigskip\bigskip

\centerline{\framebox{\textit{Phys.~Rev.} D 70 (2004) 125004}}

\bigskip\bigskip


\section{\label{sec:intro}Introduction}

The issue of Lorentz symmetry in a noncommutative field theory has been debated \cite{CHKLO, BGGKPSW, IS, CCZ, KMOU, AV, OMI, MOU, Mor, Gho} seriously, but it still remains a challenge leading to fresh insights \cite{CKNT, CPT}. The problem stems from the fact that pointwise multiplication of operators is replaced by a star multiplication:
\begin{equation}\label{lo-1}
A(x)B(x) \to A(x) \star B(x),
\end{equation}
which is defined as\footnote{This is the so called canonical definition. There are other realizations like the Lie-algebra valued structure or the $q$-deformed structure \cite{MSSW}.}
\begin{equation}\label{star.d}
A(x)\star B(x)=\left.\exp\left(\frac{\mathrm{i}}{2}\theta^{\alpha\beta}\partial_{\alpha}\partial'_{\beta}\right)A(x)B(x')\right|_{x'=x},
\end{equation}
where $\theta^{\alpha\beta}$ is a constant anti-symmetric object. Hence the ordinarily vanishing commutators among spacetime coordinates acquire a non-trivial form:
\begin{equation}\label{xxmoyal}
[x^\mu, x^\nu] \to [x^\mu, x^\nu]_{\star} \equiv x^\mu \star x^\nu - x^\nu \star x^\mu = \mathrm{i}\theta^{\mu\nu}.
\end{equation}
Since $\theta^{\mu\nu}$ is constant, theories defined on such a noncommutative spacetime are considered to violate Lorentz invariance.

Nevertheless, in spite of this vexing problem, the basic issues of noncommutative field theory, like unitarity \cite{GM}, causality \cite{SST}, mixing of UV/IR divergences \cite{MRS}, anomalies \cite{AS, BG, Rab2} are discussed in a formally Lorentz-invariant manner, using the representaion of Poincar\'e algebra. To achieve a reconciliation, therefore, it is essential to obtain a conceptually cleaner understanding of Lorentz symmetry and its interpretaion in the noncommutative context. The aim of the present paper is precisely to provide such a study.

We adopt a Noether-like approach\footnote{A somewhat similar approach, but with a different viewpoint, was followed in Ref.~\cite{IS}.} to analyse the various spacetime symmetries of noncommutative electrodynamics. This paper deals with the classical (non-quantized) electromagnetic field. Although the present study is confined to the $\mathrm{U}$(1) group, it can be extended to other (non-Abelian) groups. Since $\theta^{\mu\nu}$ is a constant, it appears as a background field in noncommutative electrodynamics. The Noether analysis, which is usually done for dynamical variables, is reformulated to include background fields. Now there are two possibilities for a constant $\theta^{\mu\nu}$. It may either be the same constant in all frames or it may transform as a second-rank tensor, taking different constant values in different frames. It is found that although the criterion for preserving translational invariance is the same in both cases, the criterion for Lorentz invariance (invariance under rotations and boosts) is different. An explicit computation shows that the criterion for Lorentz symmetry is satisfied only when $\theta^{\mu\nu}$ transforms as a tensor. Translational invariance is always satisfied. We also show that the transformations are dynamically consistent since the Noether charges correctly generate the transformations of an arbitrary function of canonical variables. Also, these charges satisfy the appropriate Lie brackets among themselves.

As is well known, noncommutative electrodynamics can be studied in two formulations; either in terms of the original noncommutative variables or, alternatively, in terms of its commutative equivalents obtained by using the Seiberg--Witten maps \cite{SW}. Our analysis has been carried out in both formulations, up to first order in $\theta$. A complete equivalence among the results has also been established. This is rather non-trivial since there are examples where this equivalence does not hold. For example, the IR problem found in noncommutative field theory \cite{MST, CR} is absent in the commutative-variable approach \cite{BGPSW}, revealing an inequivalence, at least on a perturbative level.

It is reassuring to note that an important feature \cite{AV} of quantum field theory on 4-dimensional noncommutative spacetime, namely, the invariance for a constant non-transforming $\theta$ under the $\mathrm{SO}(1,1)\times\mathrm{SO}(2)$ subgroup of Lorentz group is reproduced by the criteria found here. This has been shown in both the commutative and noncommutative descriptions.

The paper is organized as follows. In Sec.~\ref{sec:review}, the occurrence of noncommutative algebra in various approaches and their possible connections is briefly reviewed. In Sec.~\ref{sec:toy}, the implications of Lorentz symmetry in a toy model comprising a usual Maxwell field coupled to an external source are discussed. Sections \ref{sec:ncelectro-c} and \ref{sec:ncelectro-nc} provide a detailed account of Lorentz symmetry in noncommutative electrodynamics, first in the commutative-variable approach and then in terms of noncommutative variables, respectively. Concluding remarks are left for Sec.~\ref{sec:conclu}.


\section{\label{sec:review}A brief review of noncommutative algebra}

We start by briefly reviewing Snyder's algebra \cite{Sny}. The special theory of relativity may be based on the invariance of the indefinite quadratic form
\begin{equation}\label{sr}
S^{2} = (x^{0})^{2}-(x^{1})^{2}-(x^{2})^{2}-(x^{3})^{2} = -x_{\mu}x^{\mu}
\end{equation}
for transformation from one inertial frame to another. We shall use $(-, +, +, +)$ signature for the flat Minkowski metric $\eta_{\mu\nu}$. It is usually assumed that the variables $x^{\mu}$ take on a continuum of values and that they may take on these values simultaneously. Snyder considered a different situation. He considered Hermitian operators, $\op{x}^{\mu}$, for the spacetime coordinates of a particular Lorentz frame. He further assumed that the spectra of spacetime coordinate operators $\op{x}^{\mu}$ are invariant under Lorentz transformations. The later assumption is evidently satisfied by the usual spacetime continuum, however it is not the only solution. Snyder showed that there exists a Lorentz-invariant spacetime in which there is a natural unit of length.

To find operators $\op{x}^{\mu}$ possessing Lorentz-invariant spectra, Snyder considered the homogeneous quadratic form
\begin{equation}\label{qf}
-(y)^{2} = (y_{0})^{2}-(y_{1})^{2}-(y_{2})^{2}-(y_{3})^{2}-(y_{4})^{2} = -y_{\mu}y^{\mu}-(y_{4})^{2},
\end{equation}
in which $y$'s are assumed to be real variables. Now $\op{x}^{\mu}$ are defined by means of the infinitesimal elements of the group under which quadratic form \eqref{qf} is invariant. The $\op{x}^{\mu}$ are taken as
\begin{equation}\label{opx}
\op{x}^{\mu} = \mathrm{i}a\left(y_{4}\frac{\partial}{\partial y_{\mu}}-y^{\mu}\frac{\partial}{\partial y_{4}}\right),
\end{equation}
in which $a$ is the natural unit of length. These operators are assumed to be Hermitian and operate on the single-valued functions of $y_{\mu}, y_{4}$. The spectra of $\op{x}^{i}$, $i=1,2,3$, are discrete, but $\op{x}^{0}$ has a continuous spectrum extending from $-\infty$ to $+\infty$. Transformations which leave the quadratic form \eqref{qf} and $y_{4}$ invariant are covariant Lorentz transformations on the variables $y_{1}$, $y_{2}$, $y_{3}$ and $y_{0}$, and these transformations induce contravariant Lorentz transformations in $\op{x}^{\mu}$.

Now six additional operators are defined as
\begin{equation}\label{opM}
\op{M}^{\mu\nu} = -\mathrm{i}\left(y^{\mu}\frac{\partial}{\partial y_{\nu}}-y^{\nu}\frac{\partial}{\partial y_{\mu}}\right),
\end{equation}
which are the infinitesimal elements of the four-dimensional Lorentz group. The ten operators defined in Eqs.~\eqref{opx} and \eqref{opM} have the following commutation relations:
\begin{gather}
\label{xx}
\left[\op{x}^{\mu}, \op{x}^{\nu}\right] = \mathrm{i}a^{2}\op{M}^{\mu\nu},\\
\label{Mx}
\left[\op{M}^{\mu\nu}, \op{x}^{\lambda}\right] = \mathrm{i}\left(\op{x}^{\mu}\eta^{\nu\lambda}-\op{x}^{\nu}\eta^{\mu\lambda}\right),\\
\label{MM}
\left[\op{M}^{\mu\nu}, \op{M}^{\alpha\beta}\right] = \mathrm{i}\left(\op{M}^{\mu\beta}\eta^{\nu\alpha}-\op{M}^{\mu\alpha}\eta^{\nu\beta}+\op{M}^{\nu\alpha}\eta^{\mu\beta}-\op{M}^{\nu\beta}\eta^{\mu\alpha}\right).
\end{gather}
The Lorentz $\mathrm{SO}(3, 1)$ symmetry given in Eq.~\eqref{MM} is extended to $\mathrm{SO}(4, 1)$ symmetry specified by Eqs.~\eqref{xx}--\eqref{MM}.

Since the position operators $\op{x}^{i}$ have discrete spectra, we can understand it in terms of a non-zero minimal uncertainty in positions. It is possible to obtain the space part of Snyder algebra by considering the generalized Heisenberg algebra\footnote{The space part of Snyder algebra can also be obtained from another generalized Heisenberg algebra considered in Ref.~\cite{KMM}.} (with $\hbar = 1$):
\begin{equation}\label{k-xp}
\left[\op{x}_i, \op{p}_j\right] = \mathrm{i}\delta_{ij}\sqrt{1+a^{2}p_{k}p_{k}},
\end{equation}
which implies non-zero minimal uncertainties in position coordinates, and preserves the rotational symmetry. Representing the generalized Heisenberg algebra on momentum wave functions $\psi(p) = \langle p|\psi\rangle$,
\begin{gather}
\label{k-ppsi}
\op{p}_i \psi(p) = p_i \psi(p),\\
\label{k-xpsi}
\op{x}_i \psi(p) = \mathrm{i}\sqrt{1+a^{2}p_k p_k} \partial_{p_i} \psi(p),
\end{gather}
we get the commutation relation among the position operators:
\begin{equation}\label{k-xx}
\left[\op{x}_i, \op{x}_j\right] = -a^{2}\left(p_{i}\partial_{p_{j}}-p_{j}\partial_{p_{i}}\right) \equiv \mathrm{i}a^{2}\op{M}_{ij},
\end{equation}
where we have defined
\begin{equation}\label{k-M.d}
\op{M}_{ij} = \mathrm{i}\left(p_{i}\partial_{p_{j}}-p_{j}\partial_{p_{i}}\right).
\end{equation}
Thus we have
\begin{gather}
\label{k-xM}
\left[\op{M}_{ij}, \op{x}_{k}\right] = \mathrm{i}\left(\op{x}_{i}\delta_{jk}-\op{x}_{j}\delta_{ik}\right),\\
\label{k-MM}
\left[\op{M}_{ij}, \op{M}_{kl}\right] = \mathrm{i}\left(\op{M}_{il}\delta_{jk}-\op{M}_{ik}\delta_{jl}+\op{M}_{jk}\delta_{il}-\op{M}_{jl}\delta_{ik}\right).
\end{gather}
The algebra \eqref{k-xx}, \eqref{k-xM} and \eqref{k-MM} exactly reproduces the space part of the Snyder algebra \eqref{xx}--\eqref{MM}.

Doplicher et al.~\cite{DFR1, DFR2} proposed a new algebra (DFR algebra) of a noncommutative spacetime through considerations on the spacetime uncertainty relations derived from quantum mechanics and general relativity. This algebra defines a Lorentz-invariant noncommutative spacetime different from Snyder's quantized spacetime. Their algebra is given by
\begin{gather}
\label{xx-dfr}
\left[\op{x}^{\mu}, \op{x}^{\nu}\right] = \mathrm{i}\opgr{\theta}^{\mu\nu},\\
\label{thx}
\left[\opgr{\theta}^{\mu\nu}, \op{x}^{\lambda}\right] = 0,\\
\label{thth}
\left[\opgr{\theta}^{\mu\nu}, \opgr{\theta}^{\alpha\beta}\right] = 0.
\end{gather}

Recently, Carlson et al.~\cite{CCZ} rederived this DFR algebra by ``contraction'' of Snyder's algebra. For that they considered
\begin{equation}\label{Mth}
\op{M}^{\mu\nu} = \frac{1}{b}{\opgr{\theta}^{\mu\nu}},
\end{equation}
and the limits $b \to 0$, $a \to 0$ with the ratio of $a^{2}$ and $b$ held fixed: $(a^{2}/b) \to 1$.
The result of this contraction is the algebra given by Eqs.~\eqref{xx-dfr}--\eqref{thth}. It also follows that
\begin{equation}\label{M-th}
\left[\op{M}^{\mu\nu}, \opgr{\theta}^{\alpha\beta}\right] = \mathrm{i}\left(\opgr{\theta}^{\mu\beta}\eta^{\nu\alpha}+\opgr{\theta}^{\nu\alpha}\eta^{\mu\beta}-\opgr{\theta}^{\mu\alpha}\eta^{\nu\beta}-\opgr{\theta}^{\nu\beta}\eta^{\mu\alpha}\right).
\end{equation}
Since $a \to 0$ is a part of the limit, the contracted algebra corresponds to a continuum limit of Snyder's quantized spacetime.

The validity of this contraction process is questionable. Let us recall the familiar contractions of the group $\mathrm{SO}(3)$ to the group $\mathrm{E}_2$, and of the Poincar\'e group to the Galilean group. In the limit of infinite radius, $\mathrm{SO}(3)$, which is the symmetry group of the surface of the sphere, contracts to $\mathrm{E}_2$, the symmetry group of a plane. Likewise, in the low-velocity limit, the Poincar\'e group contracts to the Galilean group. These contractions involve taking limit of one parameter only whereas the above mentioned contraction of Snyder algebra to DFR algebra is achieved by taking limits of two parameters, $a \to 0$ and $b \to 0$. Furthermore, in the standard group contraction we can identify a mapping among the generators of the two groups, but in the mapping \eqref{Mth}, $\opgr{\theta}^{\mu\nu}$ is not a generator associated with any symmetry group. In this context, therefore, we agree with Kase et al.~\cite{KMOU} that there is no connection between the two algebras.

In this paper we shall consider noncommutative electrodynamics which is obtained by a standard deformation of the usual (commutative) Maxwell theory, replacing pointwise multiplication by a star multiplication defined by Eq.~\eqref{star.d}. We shall show in what precise sense Lorentz symmetry is interpreted to be valid, or otherwise. To facilitate our analysis we first develop the formulation in the context of a simple toy model.


\section{\label{sec:toy}A toy model}

We know from Noether's theorem that the invariance of action under a symmetry group, and a spacetime transformation in particular, implies the existence of a current $J^{\mu}$ satisfying a continuity equation $\partial_{\mu}J^{\mu}=0$. We shall now investigate what happens when the action contains vector or tensor parameters which are not included in the configuration space, i.e., there are external vector or tensor parameters in the theory. Before we consider the noncommutative Maxwell theory, which contains a tensor parameter $\theta^{\alpha\beta}$, it will be advantageous to first start with a simpler case.

We consider ordinary Maxwell theory with the potential coupled to an external source:
\begin{equation}\label{901}
S \equiv \int\!\mathrm{d}^{4}x\,\mathscr{L} = -\int\!\mathrm{d}^{4}x\,\left(\frac{1}{4}F_{\mu\nu}F^{\mu\nu}+j^{\mu}A_{\mu}\right).
\end{equation}
Here $j_{\mu}$ is taken to be a constant vector, i.e., it is constant but transforms as a vector when we go from one coordinate frame to another.\footnote{Later we shall also consider the case where $j^{\mu}$ does not transform like a vector but is fixed for all frames. In that case, one expects that the Lorentz invariance of the action will not be preserved.} Here we would like to mention that for the realistic current sources, $j^{\mu}$ corresponds to a vector function which is localized in space. In this sense, therefore, $j^{\mu}$ should be treated as a hypothetical source as it has been taken to be constant throughout. We are just interested in studying the Lorentz-transformation property of this system.

Let us consider an infinitesimal transformation of the coordinate system:
\begin{equation}\label{902}
x^{\mu} \rightarrow x'^{\mu} = x^{\mu}+\delta x^{\mu},
\end{equation}
under which $A^{\mu}$ and $j^{\mu}$ transform as
\begin{gather}
\label{903-1}
A^{\mu}(x) \rightarrow A'^{\mu}(x') = A^{\mu}(x)+\delta A^{\mu}(x),\\
\label{903}
j^{\mu} \rightarrow j'^{\mu} = j^{\mu}+\delta j^{\mu}.
\end{gather}
The change in the action resulting from these transformations is
\begin{equation}\label{904}
\delta S = \int_{\Omega'}\!\mathrm{d}^{4}x'\,\mathscr{L}\left(A'_{\nu}(x'),\partial'_{\mu}A'_{\nu}(x');j'_{\nu}\right)-\int_{\Omega}\!\mathrm{d}^{4}x\,\mathscr{L}\left(A_{\nu}(x),\partial_{\mu}A_{\nu}(x);j_{\nu}\right),
\end{equation}
where $\Omega$ is an arbitrarily large closed region of spacetime and $\Omega'$ being the transform of $\Omega$ under the coordinate change \eqref{902}. The above change in action can be rewritten as
\begin{equation}\label{905}
\begin{split}
\delta S &= \int_{\Omega}\!\mathrm{d}^{4}x\,\left[\mathscr{L}\left(A'_{\nu}(x),\partial_{\mu}A'_{\nu}(x);j'_{\nu}\right)-\mathscr{L}\left(A_{\nu}(x),\partial_{\mu}A_{\nu}(x);j_{\nu}\right)\right]\\
& \quad{}+\int_{\Omega'-\Omega}\!\mathrm{d}^{4}x\,\mathscr{L}\left(A'_{\nu}(x),\partial_{\mu}A'_{\nu}(x);j'_{\nu}\right).
\end{split}
\end{equation}
The last term, an integral over the infinitesimal volume $\Omega'-\Omega$, can be written as an integral over the boundary $\partial\Omega$:
\begin{equation}\label{9051}
\begin{split}
\int_{\Omega'-\Omega}\!\mathrm{d}^{4}x\,\mathscr{L}\left(A'_{\nu},\partial_{\mu}A'_{\nu};j'_{\nu}\right) &= \int_{\partial\Omega}\!\mathrm{d}S_{\lambda}\,\delta x^{\lambda}\mathscr{L}(A_{\nu},\partial_{\mu}A_{\nu};j_{\nu})\\
&= \int_{\Omega}\!\mathrm{d}^{4}x\,\partial_{\lambda}\left[\delta x^{\lambda}\mathscr{L}(A_{\nu},\partial_{\mu}A_{\nu};j_{\nu})\right],
\end{split}
\end{equation}
where Gauss theorem has been used in the last step.
For any function $f(x)$, we can write
\begin{equation}\label{906}
\delta f = f'(x')-f(x) = \delta_{0}f + \delta x^{\mu}\partial_{\mu}f,
\end{equation}
where $\delta_{0}f = f'(x)-f(x)$ is the functional change. Note that since we have taken $j^{\mu}$ to be constant, $\delta_{0}j^{\mu}=\delta j^{\mu}$. Now we have
\begin{equation}\label{911}
\begin{split}
&\mathscr{L}\left(A'_{\nu}(x),\partial_{\mu}A'_{\nu}(x);j'_{\nu}\right)-\mathscr{L}\left(A_{\nu}(x),\partial_{\mu}A_{\nu}(x);j_{\nu}\right)\\
&\quad= \frac{\partial\mathscr{L}}{\partial A_{\nu}}\delta_{0}A_{\nu}+\frac{\partial\mathscr{L}}{\partial(\partial_{\mu}A_{\nu})}\delta_{0}\partial_{\mu}A_{\nu}+\frac{\partial\mathscr{L}}{\partial j_{\nu}}\delta j_{\nu}.
\end{split}
\end{equation}
Using the equation of motion
\begin{equation}\label{eom1}
\frac{\partial \mathscr{L}}{\partial A_{\nu}}-\partial_{\mu}\left(\frac{\partial \mathscr{L}}{\partial(\partial_{\mu}A_{\nu})}\right) = 0,
\end{equation}
and the relations \eqref{9051} and \eqref{911}, we can cast Eq.~\eqref{905} as
\begin{equation*}
\delta S = \int_{\Omega}\!\mathrm{d}^{4}x\,\left[\partial_{\mu}\left(\mathscr{L}\delta x^{\mu}+\frac{\partial\mathscr{L}}{\partial(\partial_{\mu}A_{\nu})}\delta_{0}A_{\nu}\right)+\frac{\partial\mathscr{L}}{\partial j_{\nu}}\delta j_{\nu}\right].
\end{equation*}
In view of Eq.~\eqref{906}, we can write\footnote{Now onwards we drop the explicit display of $\Omega$ as we take this to correspond to entire spacetime in a suitable limit.}
\begin{equation}\label{913}
\delta S = \int\!\mathrm{d}^{4}x\,\left[\partial_{\mu}\left(\frac{\partial\mathscr{L}}{\partial(\partial_{\mu}A_{\nu})}\delta A_{\nu}-T^{\mu\nu}\delta x_{\nu}\right)+\frac{\partial\mathscr{L}}{\partial j_{\nu}}\delta j_{\nu}\right],
\end{equation}
where $T^{\mu\nu}$ is the canonical energy--momentum tensor defined by
\begin{equation}\label{914}
T^{\mu\nu} = \frac{\partial\mathscr{L}}{\partial(\partial_{\mu}A_{\sigma})}\partial^{\nu}A_{\sigma}-\eta^{\mu\nu}\mathscr{L}.
\end{equation}

For spacetime translations, $\delta x^{\mu}=a^{\mu}$, a constant, while $\delta A_{\mu}=0$ and $\delta j_{\mu}=0$. So the invariance of the action under translations implies
\begin{equation*}
\int\!\mathrm{d}^{4}x\,(\partial_{\mu}T^{\mu\nu})a_{\nu} = 0.
\end{equation*}
Since it is true for arbitrary $a_{\nu}$, we must have
\begin{equation}\label{916}
\partial_{\mu}T^{\mu\nu} = 0.
\end{equation}
This is the criterion for translational invariance of the action.

In the case of infinitesimal Lorentz transformations (rotations and boosts), $\delta x_{\mu}=\omega_{\mu\nu}x^{\nu}$, $\delta A_{\mu}=\omega_{\mu\nu}A^{\nu}$ and $\delta j_{\mu}=\omega_{\mu\nu}j^{\nu}$, where $\omega_{\mu\nu}$ is constant and anti-symmetric. So the invariance of the action implies
\begin{equation*}
\int\!\mathrm{d}^{4}x\,\left[\partial_{\mu}\bigg(\frac{\partial\mathscr{L}}{\partial(\partial_{\mu}A_{\lambda})}A^{\rho}-\frac{\partial\mathscr{L}}{\partial(\partial_{\mu}A_{\rho})}A^{\lambda}-T^{\mu\lambda}x^{\rho}+T^{\mu\rho}x^{\lambda}\bigg)+\frac{\partial \mathscr{L}}{\partial j_{\lambda}}j^{\rho}-\frac{\partial \mathscr{L}}{\partial j_{\rho}}j^{\lambda}\right]\omega_{\lambda\rho} = 0.
\end{equation*}
Since it is true for arbitrary $\omega_{\lambda\rho}$, we must have
\begin{equation}\label{918}
\partial_{\mu}M^{\mu\lambda\rho}+\frac{\partial \mathscr{L}}{\partial j_{\lambda}}j^{\rho}-\frac{\partial \mathscr{L}}{\partial j_{\rho}}j^{\lambda} = 0,
\end{equation}
where
\begin{equation}\label{9181}
M^{\mu\lambda\rho} = \frac{\partial\mathscr{L}}{\partial(\partial_{\mu}A_{\lambda})}A^{\rho}-\frac{\partial\mathscr{L}}{\partial(\partial_{\mu}A_{\rho})}A^{\lambda}-T^{\mu\lambda}x^{\rho}+T^{\mu\rho}x^{\lambda}.
\end{equation}
Therefore, the criterion for Lorentz invariance of the action is
\begin{equation}\label{918-22}
\partial_{\mu}M^{\mu\lambda\rho}-A^{\lambda}j^{\rho}+A^{\rho}j^{\lambda} = 0.
\end{equation}

Now we shall obtain the criteria for translational invariance and Lorentz invariance of the action when $j^{\mu}$ is not a genuine vector but has the same constant value in all frames. In that case we have $\delta j^{\mu} = 0$ not only under translations but also under Lorentz transformations. Therefore the last term inside the parentheses on the right-hand side of Eq.~\eqref{913} drops out and the criteria for the invariance of the action turn out to be
\begin{gather}
\label{crit1}
\partial_{\mu}T^{\mu\nu} = 0,\\
\label{crit2}
\partial_{\mu}M^{\mu\lambda\rho} = 0.
\end{gather}
Thus, the criterion for translational invariance is the same irrespective of whether $j^{\mu}$ is a genuine vector or not. However, this is not the case with the criterion for Lorentz invariance.

Now we shall explicitly evaluate $\partial_{\mu}T^{\mu\nu}$ and $\partial_{\mu}M^{\mu\lambda\rho}$ for our toy model \eqref{901}. This will obviously be independent of whether $j^{\mu}$ transforms like a vector or not. Using
\[
\partial^{\nu}\mathscr{L} = \frac{\partial \mathscr{L}}{\partial A_{\rho}}\partial^{\nu}A_{\rho}+\frac{\partial \mathscr{L}}{\partial(\partial_{\kappa} A_{\rho})}\partial^{\nu}\partial_{\kappa}A_{\rho},
\]
the equation of motion \eqref{eom1}, and the definition \eqref{914} of energy--momentum tensor, we find
\begin{equation}\label{find1}
\partial_{\mu}T^{\mu\nu} = 0.
\end{equation}
Also, using the equation of motion \eqref{eom1}, Eq.~\eqref{find1} and the defintion \eqref{9181} of $M^{\mu\lambda\rho}$, we find for our theory \eqref{901} that
\begin{equation}\label{dM}
\partial_{\mu}M^{\mu\lambda\rho} = A^{\lambda}j^{\rho}-A^{\rho}j^{\lambda}.
\end{equation}
As mentioned earlier, the results \eqref{find1} and \eqref{dM} do not depend whether $j^{\mu}$ transforms like a vector or not.

We have seen that the criterion for translational invariance is the same,  $\partial_{\mu}T^{\mu\nu} = 0$, in both the cases, independent of whether $j^{\mu}$ transforms like a vector or not. This is satisfied in view of Eq.~\eqref{find1}, thereby indicating that our toy model has translational invariance in both the cases. However, the criterion for Lorentz invariance is different in the two cases---see Eqs.~\eqref{918-22} and \eqref{crit2}---whereas what we have actually found is given by Eq.~\eqref{dM}. Since this agrees with the criterion \eqref{918-22}, our model has Lorentz invariance only when $j^{\mu}$ transforms like a vector, and not in the other case.

We shall now show that using the Noether charges
\begin{equation}\label{gen}
P^{\mu} = \int\!\mathrm{d}^{3}x\,T^{0\mu}, \quad J^{\mu\nu} = \int\!\mathrm{d}^{3}x\,M^{0\mu\nu},
\end{equation}
and the canonical equal-time Poisson brackets $\{A_{\mu}(t, \mathbf{x}), \pi^{\nu}(t, \mathbf{y})\} = \delta_{\mu}^{\nu}\delta^{3}(\mathbf{x}-\mathbf{y})$, we can generate the transformations of the dynamical variables $A_i$ and $\pi^{i}$:
\begin{equation}\label{AQL}
\left\{A_{i}, Q_{V}\right\} = \mathcal{L}_{V}A_{i}, \quad \left\{\pi^{i}, Q_{V}\right\} = \mathcal{L}_{V}\pi^{i},
\end{equation}
where $Q_{\partial_{\mu}}=P_{\mu}$, $Q_{x_{[\mu}\partial_{\nu]}}=J_{\mu\nu}$ and  $\mathcal{L}_{V}A_{i}$ stands for the Lie derivative\footnote{If $W{}^{\alpha\ldots\beta}_{\mu\ldots\nu}(x) \to W'{}^{\alpha\ldots\beta}_{\mu\ldots\nu}(x')$ for an arbitrary tensor field under the infinitesimal transformation $x^{\mu} \to x'^{\mu} = x^{\mu}-bV^{\mu}$, then the Lie derivative of $W(x)$ with respect to the vector field  $V(x) = V^{\mu}(x)\partial_{\mu}$ is defined as
\[
\left(\mathcal{L}_{V}W\right)^{\alpha\ldots\beta}_{\mu\ldots\nu}(x) = \lim_{b \to 0}\frac{1}{b}\left(W'{}^{\alpha\ldots\beta}_{\mu\ldots\nu}(x)-W{}^{\alpha\ldots\beta}_{\mu\ldots\nu}(x)\right).
\]
} of the field $A_i$ with respect to the vector field $V$ associated with the charge $Q_{V}$.

The canonical momenta of the theory are
\begin{gather}
\label{pi0}
\pi^{0} = \frac{\partial\mathscr{L}}{\partial(\partial_{0}A_{0})} = 0,\\
\label{pii}
\pi^{i} = \frac{\partial\mathscr{L}}{\partial(\partial_{0}A_{i})} = F^{i0}.
\end{gather}
It follows from the definitions \eqref{914} and \eqref{9181} that
\begin{gather}
\label{T00}
T^{00} = \pi^{i}\partial_{i}A^{0}-\frac{1}{2}\pi^{i}\pi_{i}-\frac{1}{4}F_{ij}F^{ij}-j^{\mu}A_{\mu},\\
\label{T0i}
T^{0i} = \pi^{j}\partial^{i}A_{j},\\
\label{M00i}
M^{00i} = -T^{00}x^{i}-\pi^{i}A^{0}+x^{0}\pi^{j}\partial^{i}A_{j},\\
\label{M0ij}
M^{0ij} = \pi^{i}A^{j}-x^{j}\pi^{k}\partial^{i}A_{k}-\pi^{j}A^{i}+x^{i}\pi^{k}\partial^{j}A_{k},
\end{gather}
where we have used Eq.~\eqref{pii} to eliminate velocities in favour of momenta. Now we compute the Poisson brackets of the field $A_{i}$ with the charges:
\begin{gather}
\label{AP1}
\left\{A_{i}, P_{j}\right\} = \partial_{j}A_{i},\\
\label{AP2}
\left\{A_{i}, P_{0}\right\} = \partial_{i}A_{0}+\pi_{i} = \partial_{0}A_{i},\\
\label{AJ1}
\left\{A_{i}, J_{kl}\right\} = \eta_{ik}A_{l}-x_{l}\partial_{k}A_{i}-\eta_{il}A_{k}+x_{k}\partial_{l}A_{i},\\
\label{AJ2}
\left\{A_{i}, J_{0l}\right\} = -x_{l}\left(\partial_{i}A_{0}+\pi_{i}\right)-\eta_{il}A_{0}+x_{0}\partial_{l}A_{i} = -x_{l}\partial_{0}A_{i}-\eta_{il}A_{0}+x_{0}\partial_{l}A_{i},
\end{gather}
where the definition \eqref{pii} of momenta has been used in the second steps of Eqs.~\eqref{AP2} and \eqref{AJ2}.
Since
\begin{gather}
\label{ld1}
\mathcal{L}_{\partial_{\mu}}A_{i} = \partial_{\mu}A_{i},\\
\label{ld2}
\mathcal{L}_{x_{[\mu}\partial_{\nu]}}A_{i} = \eta_{i\mu}A_{\nu}-x_{\nu}\partial_{\mu}A_{i}-\eta_{i\nu}A_{\mu}+x_{\mu}\partial_{\nu}A_{i},
\end{gather}
it follows that
\begin{equation}
\label{APL}
\left\{A_{i}, P_{\mu}\right\} = \mathcal{L}_{\partial_{\mu}}A_{i}, \quad
\left\{A_{i}, J_{\mu\nu}\right\} = \mathcal{L}_{x_{[\mu}\partial_{\nu]}}A_{i}.
\end{equation}
The brackets of the momenta $\pi_{i}$ with the charges are
\begin{gather}
\label{piP1}
\left\{\pi_{i}, P_{j}\right\} = \partial_{j}\pi_{i},\\
\label{piP2}
\left\{\pi_{i}, P_{0}\right\} = \partial_{k}{F^{k}}_{i}-j_{i} = \partial_{0}\pi_{i},\\
\label{piJ1}
\left\{\pi_{i}, J_{kl}\right\} = \eta_{ik}\pi_{l}-x_{l}\partial_{k}\pi_{i}-\eta_{il}\pi_{k}+x_{k}\partial_{l}\pi_{i},\\
\label{piJ2}
\left\{\pi_{i}, J_{0l}\right\} = -x_{l}\left(\partial_{k}{F^{k}}_{i}-j_{i}\right)+x_{0}\partial_{l}\pi_{i}-F_{li} = -x_{l}\partial_{0}\pi_{i}+x_{0}\partial_{l}\pi_{i}-F_{li},
\end{gather}
where, in the second steps of Eqs.~\eqref{piP2} and \eqref{piJ2}, we have used $\partial_{0}\pi^{i}=\partial_{k}F^{ki}-j^{i}$ which is a consequence of the equation of motion \eqref{eom1}:
\begin{gather}
\label{motion}
\partial_{\mu}F^{\mu\nu}-j^{\nu}=0\\
\nonumber
\Rightarrow \quad \partial_{0}F^{0i}+\partial_{k}F^{ki}-j^{i} = -\partial_{0}\pi^{i}+\partial_{k}F^{ki}-j^{i}=0.
\end{gather}
Since\footnote{It is perhaps worthwhile to mention that while computing the Lie derivative of $\pi^{i}$, one should keep in mind that $\pi^{i}$ are not the components of a 4-vector. Rather, $\pi^{i}$ are the components of a tensor, $\pi^{i}=F^{i0}$.}
\begin{gather}
\label{pild1}
\mathcal{L}_{\partial_{\mu}}\pi_{i} = \partial_{\mu}\pi_{i},\\
\label{pild2}
\mathcal{L}_{x_{[k}\partial_{l]}}\pi_{i} = \eta_{ik}\pi_{l}-x_{l}\partial_{k}\pi_{i}-\eta_{il}\pi_{k}+x_{k}\partial_{l}\pi_{i},\\
\label{pild3}
\mathcal{L}_{(x_{0}\partial_{l}-x_{l}\partial_{0})}\pi_{i} = -x_{l}\partial_{0}\pi_{i}+x_{0}\partial_{l}\pi_{i}-F_{li},
\end{gather}
it follows that
\begin{equation}
\label{piPL}
\left\{\pi_{i}, P_{\mu}\right\} = \mathcal{L}_{\partial_{\mu}}\pi_{i}, \quad
\left\{\pi_{i}, J_{\mu\nu}\right\} = \mathcal{L}_{x_{[\mu}\partial_{\nu]}}\pi_{i}.
\end{equation}
Hence we have shown that Eq.~\eqref{AQL} is indeed satisfied.

We also find that
\begin{gather}
\label{PP}
\left\{P_{i}, P_{j}\right\} = 0,\\
\label{PJ}
\left\{P_{i}, J_{kl}\right\} = \eta_{ik}P_{l}-\eta_{il}P_{k},\\
\label{JJ}
\left\{J_{ij}, J_{kl}\right\} = \eta_{jk}J_{il}+\eta_{il}J_{jk}-\eta_{ik}J_{jl}-\eta_{jl}J_{ik}.
\end{gather}
Now it follows that restricting to kinematical generators ($P_{i}$ and $J_{ij}$) only, we have
\begin{equation}\label{QQ}
\left\{Q_{U}, Q_{V}\right\} = Q_{[U,V]}.
\end{equation}
Thus we see that although $\partial_{0}Q_{V} \neq 0$ (for rotations and boosts), we still have Eqs.~\eqref{AQL} and \eqref{QQ}. This is necessary for establishing the dynamical consistency of the transformations.

It should be stressed that the Hamiltonian approach violates manifest Lorentz invariance. The fact that it gets restored is thus quite non-trivial. A possible way to see the manifest violation is through Eq.~\eqref{pi0}. Within the Hamiltonian formulation, however, this equation really is a primary constraint and the equality is only ``weakly'' valid \cite{Dirac}. Time-conserving the primary constraint leads to a secondary (Gauss) constraint. This is basically the zero-component of the equation of motion \eqref{motion}, expressed in phase-space variables:
\begin{equation}\label{gauss1}
\partial_{i}\pi^{i}-j^{0} \approx 0.
\end{equation}
There are no further constraints. These constraints do not affect the realization of the 3D Euclidean symmetry \eqref{PP}--\eqref{JJ}.


\section{\label{sec:ncelectro-c}Noncommutative electrodynamics: commutative-variable approach}

We now generalize the case of vector source considered in the previous section to anti-symmetric tensor ``source'' $\theta^{\mu\nu}$. We take the noncommutative Maxwell theory:
\begin{equation}\label{l-105}
\nc{S}=-\frac{1}{4}\int\!\mathrm{d}^{4}x\,\left(\nc{F}_{\mu\nu}\star\nc{F}^{\mu\nu}\right),
\end{equation}
where the noncommutative field strength is
\begin{equation}\label{l-106}
\nc{F}_{\mu\nu}=\partial_{\mu}\nc{A}_{\nu}-\partial_{\nu}\nc{A}_{\mu}-\mathrm{i}\left[\nc{A}_{\mu}, \nc{A}_{\nu}\right]_{\star}.
\end{equation}
On applying the Seiberg--Witten maps~\cite{SW},
\begin{gather}
\label{101-lo}
\nc{A}_{\mu} = A_{\mu}-\frac{1}{2}\theta^{\alpha\beta}A_{\alpha}\left(\partial_{\beta}A_{\mu}+F_{\beta\mu}\right)+O(\theta^{2}),\\
\label{107-lo}
\nc{F}_{\mu\nu} = F_{\mu\nu}-\theta^{\alpha\beta}\left(A_{\alpha}\partial_{\beta}F_{\mu\nu}+F_{\mu\alpha}F_{\beta\nu}\right)+O(\theta^{2}),
\end{gather}
we get the effective theory in terms of usual (commutative) variables:
\begin{equation}\label{921}
S = -\int\!\mathrm{d}^{4}x\,\left[\frac{1}{4}F_{\mu\nu}F^{\mu\nu}+\theta^{\alpha\beta}\left(\frac{1}{2}F_{\mu\alpha}F_{\nu\beta}+\frac{1}{8}F_{\beta\alpha}F_{\mu\nu}\right)F^{\mu\nu}\right]+O(\theta^2),
\end{equation}
where a boundary term has been dropped in order to express it solely in terms of the field strength. Although we have kept only linear terms in $\theta$, our conclusions are expected to hold for the full theory. The Euler--Lagrange equation of motion for this theory (in view of the fact that $\mathscr{L}$ does not have explicit dependence on $A_{\mu}$) is 
\begin{equation}\label{eomm}
\partial_{\rho}\left(\frac{\partial\mathscr{L}}{\partial(\partial_{\sigma}A_{\rho})}\right) = 0.
\end{equation}

Popular noncommutative spacetime is characterized by a constant and fixed (same value in all frames) noncommutativity parameter but here first we take $\theta^{\alpha\beta}$ to be a constant tensor parameter, i.e., it is constant but transforms as a tensor under Poincar\'e transfomations. Proceeding as in the previous section, we find that for spacetime translations, invariance of the action implies, as before,
\begin{equation}\label{916-2}
\partial_{\mu}T^{\mu\nu} = 0,
\end{equation}
with $T^{\mu\nu}$ defined as in \eqref{914}, i.e.,
\begin{equation}\label{914'}
T^{\mu\nu} = \frac{\partial\mathscr{L}}{\partial(\partial_{\mu}A_{\sigma})}\partial^{\nu}A_{\sigma}-\eta^{\mu\nu}\mathscr{L}.
\end{equation}

In case of infinitesimal Lorentz transformations, $\delta x_{\mu}=\omega_{\mu\nu}x^{\nu}$, $\delta A_{\mu}=\omega_{\mu\nu}A^{\nu}$ and $\delta \theta_{\mu\nu}=\omega_{\mu\alpha}{\theta^{\alpha}}_{\nu}-\omega_{\nu\alpha}{\theta^{\alpha}}_{\mu}$. With $M^{\mu\lambda\rho}$ defined as in \eqref{9181},
\begin{equation}\label{9181'}
M^{\mu\lambda\rho} = \frac{\partial\mathscr{L}}{\partial(\partial_{\mu}A_{\lambda})}A^{\rho}-\frac{\partial\mathscr{L}}{\partial(\partial_{\mu}A_{\rho})}A^{\lambda}-T^{\mu\lambda}x^{\rho}+T^{\mu\rho}x^{\lambda},
\end{equation}
the analogue of Eq.~\eqref{918} turns out to be
\begin{equation}\label{918-2}
\partial_{\mu}M^{\mu\lambda\rho}+2\frac{\partial \mathscr{L}}{\partial \theta_{\alpha\rho}}{\theta^{\lambda}}_{\alpha}-2\frac{\partial \mathscr{L}}{\partial \theta_{\alpha\lambda}}{\theta^{\rho}}_{\alpha} = 0,
\end{equation}
which, upon substituting
\begin{equation*}
\frac{\partial\mathscr{L}}{\partial \theta_{\alpha\rho}} = -\frac{1}{2}\left(F^{\mu\alpha}F^{\nu\rho}+\frac{1}{4}F^{\rho\alpha}F^{\mu\nu}\right)F_{\mu\nu},
\end{equation*}
gives us the criterion for Lorentz invariance of the action as
\begin{equation}\label{407n}
\partial_{\mu}M^{\mu\lambda\rho} - {\theta^{\lambda}}_{\alpha}F_{\mu\nu}\left(F^{\mu\alpha}F^{\nu\rho}+\frac{1}{4}F^{\mu\nu}F^{\rho\alpha}\right)+{\theta^{\rho}}_{\alpha}F_{\mu\nu}\left(F^{\mu\alpha}F^{\nu\lambda}+\frac{1}{4}F^{\mu\nu}F^{\lambda\alpha}\right) = 0.
\end{equation}

In the case when $\theta^{\mu\nu}$ does not transform like a tensor but is fixed in all frames, we have $\delta \theta_{\mu\nu} = 0$ under translations and Lorentz transformations. In that case, the criteria for the invariance of the action turn out to be
\begin{gather}
\label{crit1n}
\partial_{\mu}T^{\mu\nu} = 0,\\
\label{crit2n}
\partial_{\mu}M^{\mu\lambda\rho} = 0,
\end{gather}
which are the exact analogues of the criteria \eqref{crit1} and \eqref{crit2}.

Now we shall explicitly evaluate $\partial_{\mu}T^{\mu\nu}$ and $\partial_{\mu}M^{\mu\lambda\rho}$ for our model \eqref{921}. We have
\begin{equation}\label{1191}
\frac{\partial \mathscr{L}}{\partial(\partial_{\sigma}A_{\rho})} = F^{\rho\sigma}+\theta^{\alpha\sigma}F^{\mu\rho}F_{\mu\alpha}-\theta^{\alpha\rho}F^{\mu\sigma}F_{\mu\alpha}-\frac{1}{4}\theta^{\rho\sigma}F^{\mu\nu}F_{\mu\nu}+\theta^{\alpha\beta}\left({F^{\rho}}_{\alpha}{F^{\sigma}}_{\beta}+\frac{1}{2}F_{\beta\alpha}F^{\rho\sigma}\right).
\end{equation}
Taking the derivative of Eq.~\eqref{914'} and using the equation of motion \eqref{eomm}, yields
\begin{equation}\label{find1n}
\partial_{\mu}T^{\mu\nu} = 0.
\end{equation}
Similarly, taking the derivative of Eq.~\eqref{9181'}, using Eqs.~\eqref{eomm} and \eqref{find1n}, and finally substituting \eqref{1191}, we find
\begin{equation}\label{find2n}
\partial_{\mu}M^{\mu\lambda\rho} = {\theta^{\lambda}}_{\alpha}F_{\mu\nu}\left(F^{\mu\alpha}F^{\nu\rho}+\frac{1}{4}F^{\mu\nu}F^{\rho\alpha}\right)-{\theta^{\rho}}_{\alpha}F_{\mu\nu}\left(F^{\mu\alpha}F^{\nu\lambda}+\frac{1}{4}F^{\mu\nu}F^{\lambda\alpha}\right).
\end{equation}
The results \eqref{find1n} and \eqref{find2n} do not depend on whether $\theta^{\mu\nu}$ transforms like a tensor or not.

We have seen that the criterion for translational invariance is the same,  $\partial_{\mu}T^{\mu\nu} = 0$, in both the cases when $\theta^{\mu\nu}$ transforms like a tensor and when it does not. This is satisfied in view of Eq.~\eqref{find1n}. However, the criterion for Lorentz invariance is different in the two cases---see Eqs.~\eqref{407n} and \eqref{crit2n}---and what we have actually found is given by Eq.~\eqref{find2n}. Therefore, as expected, our theory has Lorentz invariance only when $\theta^{\mu\nu}$ transforms like a tensor, and not in the other case. The Seiberg--Witten maps \eqref{101-lo} and \eqref{107-lo} have an explicit Lorentz-invariant form provided that $\theta$ transforms like a Lorentz tensor, in accordance with the result found here.

As in the toy model, we now show that the Poisson bracket of the dynamical fields $A_i$ and $\pi^i$ with the charge is equal to the Lie derivative of the field with respect to the vector field associated with the charge. As usual, the Hamiltonian formulation \cite{Rab} is commenced by computing the canonical momenta of the theory:
\begin{gather}
\label{pi0n}
\pi^{0} = 0,\\
\label{piin}
\begin{split}
\pi^{i} &= F^{i0}-\theta^{mn}\left(F{^{i}}_{n}F{^{0}}_{m}+\frac{1}{2}F_{nm}F^{0i}\right)-\theta^{in}F_{kn}F^{0k}-\theta^{0n}\left(F^{0i}F_{0n}+F^{mi}F_{mn}\right)\\&\quad{}+\theta^{0i}\left(\frac{1}{4}F^{mn}F_{mn}-\frac{1}{2}F^{0m}F_{0m}\right).
\end{split}
\end{gather}
As before, Eq.~\eqref{pi0n} is interpreted as a primary constraint. Since the definition \eqref{piin} of momenta $\pi^{i}$ contains terms quadratic in ``velocities'', it is highly non-trivial to invert this relation to express velocities in terms of phase-space variables. Therefore, we now implement the condition\footnote{The simplifications achieved by this condition are well known in the Hamiltonian formulation of noncommutative gauge theories. It eliminates the higher-order time-derivatives so that the standard Hamiltonian prescription can be adopted.} $\theta^{0i}=0$,  which enables us to write down the velocities in terms of phase-space variables:
\begin{equation}\label{piin2}
F^{i0}=\pi^{i}-\theta^{mn}\left({F^{i}}_{n}\pi_{m}+\frac{1}{2}F_{nm}\pi^{i}\right)-\theta^{in}F_{kn}\pi_{k}.
\end{equation}

It follows from the definitions of $T^{\mu\nu}$ \eqref{914'} and $M^{\mu\lambda\rho}$ \eqref{9181'} that
\begin{gather}
\label{T00n}
\begin{split}
T^{00} &= \pi^{i}\partial_{i}A^{0}-\frac{1}{2}\pi^{i}\pi_{i}-\frac{1}{4}F_{ij}F^{ij}\\
& \quad{}-\theta^{ij}\left(\frac{1}{2}F_{ki}F_{mj}F^{km}+\frac{1}{8}F_{ji}F_{km}F^{km}-\frac{1}{4}F_{ji}\pi_{k}\pi^{k}-F_{kj}\pi_{i}\pi^{k}\right),
\end{split}\\
\label{T0in}
T^{0i} = \pi^{j}\partial^{i}A_{j},\\
\label{M00in}
M^{00i} = -T^{00}x^{i}-\pi^{i}A^{0}+x^{0}\pi^{j}\partial^{i}A_{j},\\
\label{M0ijn}
M^{0ij} = \pi^{i}A^{j}-x^{j}\pi^{k}\partial^{i}A_{k}-\pi^{j}A^{i}+x^{i}\pi^{k}\partial^{j}A_{k},
\end{gather}
where we have used Eq.~\eqref{piin2} to eliminate velocities in favour of momenta. Time-conserving the primary constraint with the Hamiltonian $\int \mathrm{d}^{3}x\,{T^{0}}_{0}$ yields the Gauss constraint
\begin{equation}\label{gauss2}
\partial_{i}\pi^{i} \approx 0.
\end{equation}
There are no further constraints.

Now we find
\begin{gather}
\label{AP1n}
\left\{A_{i}, P_{j}\right\} = \partial_{j}A_{i},\\
\label{AP2n}
\left\{A_{i}, P_{0}\right\} = \partial_{i}A_{0}+\pi_{i}-{\theta_{i}}^{n}F_{mn}\pi^{m}-\theta^{mn}\left(F_{in}\pi_{m}+\frac{1}{2}F_{nm}\pi_{i}\right),
\\
\label{AJ1n}
\left\{A_{i}, J_{kl}\right\} = \eta_{ik}A_{l}-\partial_{k}A_{i}x_{l}-\eta_{il}A_{k}+\partial_{l}A_{i}x_{k},\\
\label{AJ2n}
\begin{split}
\left\{A_{i}, J_{0k}\right\} &= -x_{k}\left[\partial_{i}A_{0}+\pi_{i}-{\theta_{i}}^{n}F_{mn}\pi^{m}-\theta^{mn}\left(F_{in}\pi_{m}+\frac{1}{2}F_{nm}\pi_{i}\right)\right]
+\partial_{k}A_{i}x_{0}\\&\quad{}-\eta_{ik}A_{0}.
\end{split}
\end{gather}
As in the toy model, here also we obtain
\begin{equation}\label{AQLn}
\left\{A_{i}, Q_{V}\right\} = \mathcal{L}_{V}A_{i}, \quad \left\{\pi^{i}, Q_{V}\right\} = \mathcal{L}_{V}\pi^{i}.
\end{equation}
We find that algebra \eqref{PP}--\eqref{JJ} is satisfied here also, which in turn implies that the condition \eqref{QQ} holds, i.e., restricting to $P_{i}$ and $J_{ij}$, we have
\begin{equation}\label{QQn}
\left\{Q_{U}, Q_{V}\right\} = Q_{[U,V]}.
\end{equation}

Finally, we would like to mention that there are certain choices of constant non-transforming $\theta$ for which the Lorentz invariance can be partially restored. Let us get back to Eq.~\eqref{xxmoyal} which characterizes the noncommutativity. Under Lorentz transformation, $\delta x^{\mu} = {\omega^{\mu}}_{\lambda}x^{\lambda}$, this equation imposes the following restriction on non-transforming $\theta$: 
\begin{equation}\label{lo-111}
\Omega^{\mu\nu} \equiv {\omega^{\mu}}_{\lambda}\theta^{\lambda\nu}-{\omega^{\nu}}_{\lambda}\theta^{\lambda\mu} = 0.
\end{equation}
There is no non-trivial solution of this set of equations. However, some subsets of this set of equations are soluble. It can be easily seen that the equation
\begin{equation*}
\Omega^{01} \equiv {\omega^{0}}_{2}\theta^{21}+{\omega^{0}}_{3}\theta^{31}-{\omega^{1}}_{2}\theta^{20}-{\omega^{1}}_{3}\theta^{30} = 0
\end{equation*}
is satisfied for $\theta^{02}=\theta^{03}=\theta^{12}=\theta^{13}=0$. This choice of $\theta$ also solves $\Omega^{23}=0$. Thus, invariance under a rotation in $23$-plane and under a boost in $1$-direction can be restored (for non-transforming $\theta$) by choosing
\begin{equation}\label{lo-113}
\left\{\theta^{\mu\nu}\right\} = \begin{pmatrix}0&\theta_{e}&0&0\\-\theta_{e}&0&0&0\\0&0&0&\theta_{m}\\0&0&-\theta_{m}&0
\end{pmatrix}.
\end{equation}
Likewise it can be seen that the invariance under a rotation in $13$-plane and under a boost in $2$-direction is restored for $\theta^{01}=\theta^{03}=\theta^{12}=\theta^{23}=0$, whereas for $\theta^{01}=\theta^{02}=\theta^{13}=\theta^{23}=0$, the invariance under a rotation in $12$-plane and under a boost in $3$-direction is restored. The spacetime symmetry group for these choices of $\theta$ is $[\mathrm{SO(1,1)}\times\mathrm{SO(2)}]\rtimes \mathrm{T}_{4}$, where $\rtimes$ represents semi-direct product.

We now show that these results also follow from our analysis. We have shown that the criterion for Lorentz invariance when $\theta$ does not transform is $\partial_{\mu}M^{\mu\lambda\rho}=0$, Eq.~\eqref{crit2n}. For the choice \eqref{lo-113} of $\theta$, Eq.~\eqref{find2n} indeed gives $\partial_{\mu}M^{\mu 23}=0$ and $\partial_{\mu}M^{\mu 01}=0$. Similarly, our analysis gives consistent results for the other choices of $\theta$. It is worthwhile to mention that the choice \eqref{lo-113} has recently been studied \cite{AV, Mor} and CPT theorem in noncommutative field theories has been proved~\cite{AV}.

Noncommutative gauge theories in two dimensions are always Lorentz invariant, since, in two dimensions, the noncommutativity parameter becomes proportional to the anti-symmetric tensor $\varepsilon^{\mu\nu}$, which has the same value in all frames. Our analysis is also consistent with this fact; in two dimensions, Eq.~\eqref{find2n} gives $\partial_{\mu}M^{\mu 01}=0$.


\section{\label{sec:ncelectro-nc}Noncommutative electrodynamics: noncommutative-variable approach}

Here we shall reconsider the analysis of the previous section, but in noncommutative variables. However, as in the previous section, we again restrict ourselves to the first order in $\theta$. In this approximation, the original theory~\eqref{l-105} reads
\begin{equation}\label{nl-105}
\nc{S} = -\frac{1}{2}\int\!\mathrm{d}^4x\,\left[\partial_{\mu}\nc{A}_{\nu}\left(\partial^{\mu}\nc{A}^{\nu}-\partial^{\nu}\nc{A}^{\mu}\right)+2\theta^{\alpha\beta}\partial_{\alpha}\nc{A}^{\mu}\partial_{\beta}\nc{A}^{\nu}\partial_{\mu}\nc{A}_{\nu}\right].
\end{equation}
The change of $\nc{A}_{\mu}$ under Poincar\'e transformation is dictated by the noncommutativity parameter $\theta^{\mu\nu}$ through the Seiberg--Witten map \eqref{101-lo}; $\nc{A}_{\mu}$ will transform differently depending on whether $\theta^{\mu\nu}$ transforms like a tensor or not. For spacetime translations, however, it does not matter; $\delta A_{\mu}=0$ and $\delta \theta^{\mu\nu}=0$ imply $\delta \nc{A}_{\mu}=0$. Under Lorentz transformation, $\delta A_{\mu}=\omega_{\mu\nu}A^{\nu}$, $\delta F_{\beta\mu}=\omega_{\beta\lambda}{F^{\lambda}}_{\mu}-\omega_{\mu\lambda}{F^{\lambda}}_{\beta}$, and $\delta \theta_{\mu\nu}=\omega_{\mu\alpha}{\theta^{\alpha}}_{\nu}-\omega_{\nu\alpha}{\theta^{\alpha}}_{\mu}$ if $\theta_{\mu\nu}$ transforms as a tensor, otherwise $\delta \theta_{\mu\nu}=0$ if it does not transform. Therefore, for transforming $\theta$,
 map \eqref{101-lo} gives
\begin{equation}\label{delA1-lo}
\delta \nc{A}_{\mu}=\omega_{\mu\lambda}\nc{A}^{\lambda},
\end{equation}
which is the expected noncommutative deformation of the standard transformation for a covariant vector. For non-transforming $\theta$,
\begin{equation}\label{delA2-lo}
\delta \nc{A}_{\mu} = \omega_{\mu\lambda}\nc{A}^{\lambda}-\frac{1}{2}\theta^{\alpha\beta}\omega_{\beta\lambda}\left[\nc{A}^{\lambda}\partial_{\mu}\nc{A}_{\alpha}-\nc{A}_{\alpha}\partial_{\mu}\nc{A}^{\lambda}-2\left(\nc{A}^{\lambda}\partial_{\alpha}\nc{A}_{\mu}-\nc{A}_{\alpha}\partial^{\lambda}\nc{A}_{\mu}\right)\right].
\end{equation}

Proceeding as in the case of toy model, we find that the change in action under spacetime transformations is given by
\begin{equation}\label{913-lo}
\delta \nc{S} = \int\!\mathrm{d}^{4}x\,\left[\partial_{\mu}\left(\frac{\partial\nc{\mathscr{L}}}{\partial(\partial_{\mu}\nc{A}_{\nu})}\delta \nc{A}_{\nu}-\nc{T}^{\mu\nu}\delta x_{\nu}\right)+\frac{\partial\nc{\mathscr{L}}}{\partial \theta^{\mu\nu}}\delta \theta^{\mu\nu}\right],
\end{equation}
where the canonical energy--momentum tensor is defined as
\begin{equation}\label{914-lo}
\nc{T}^{\mu\nu} = \frac{\partial\nc{\mathscr{L}}}{\partial(\partial_{\mu}\nc{A}_{\sigma})}\partial^{\nu}\nc{A}_{\sigma}-\eta^{\mu\nu}\nc{\mathscr{L}}.
\end{equation}

Therefore, the criterion for translational invariance of the action, irrespective of whether $\theta$ is a tensor or not, is
\begin{equation}\label{916-lo}
\partial_{\mu}\nc{T}^{\mu\nu} = 0,
\end{equation}
since $\delta A_{\nu} = \delta \theta^{\mu\nu} = 0$. It follows from the definition \eqref{914-lo} that the criterion \eqref{916-lo} is indeed satisfied once we use the equation of motion (Lagrangian density does not have explicit dependence on $\nc{A}_{\mu}$)
\begin{equation}\label{eom1-lo}
\partial_{\mu}\left(\frac{\partial \nc{\mathscr{L}}}{\partial(\partial_{\mu}\nc{A}_{\nu})}\right) = 0.
\end{equation}
This implies that the action~\eqref{nl-105} is invariant under translations.

In the case of transforming $\theta$, the criterion of Lorentz invariance, using the transformation \eqref{delA1-lo}, turns out to be
\begin{equation}\label{crit2nc}
\partial_{\mu}\nc{M}^{\mu\lambda\rho}-\left(\partial_{\mu}\nc{A}_{\nu}-\partial_{\nu}\nc{A}_{\mu}\right)\left({\theta^{\lambda}}_{\alpha}\partial^{\alpha}\nc{A}^{\mu}\partial^{\rho}\nc{A}^{\nu}-{\theta^{\rho}}_{\alpha}\partial^{\alpha}\nc{A}^{\mu}\partial^{\lambda}\nc{A}^{\nu}\right) = 0,
\end{equation}
where
\begin{equation}\label{9181-lo}
\nc{M}^{\mu\lambda\rho} = \frac{\partial\nc{\mathscr{L}}}{\partial(\partial_{\mu}\nc{A}_{\lambda})}\nc{A}^{\rho}-\frac{\partial\nc{\mathscr{L}}}{\partial(\partial_{\mu}\nc{A}_{\rho})}\nc{A}^{\lambda}-\nc{T}^{\mu\lambda}x^{\rho}+\nc{T}^{\mu\rho}x^{\lambda}.
\end{equation}
On the other hand, using the transformation \eqref{delA2-lo} for non-transforming $\theta$, the invariance of the action under Lorentz transformations demands
\begin{equation}\label{crit22nc}
\begin{split}
\partial_{\mu}\nc{M}^{\mu\lambda\rho}-\frac{1}{2}{\theta^{\lambda}}_{\alpha}\left(\partial_{\mu}\nc{A}_{\nu}-\partial_{\nu}\nc{A}_{\mu}\right)\partial^{\mu}\left[\nc{A}^{\alpha}(2\partial^{\rho}\nc{A}^{\nu}-\partial^{\nu}\nc{A}^{\rho})-\nc{A}^{\rho}\left(2\partial^{\alpha}\nc{A}^{\nu}-\partial^{\nu}\nc{A}^{\alpha}\right)\right]
\\+\frac{1}{2}{\theta^{\rho}}_{\alpha}\left(\partial_{\mu}\nc{A}_{\nu}-\partial_{\nu}\nc{A}_{\mu}\right)\partial^{\mu}\left[\nc{A}^{\alpha}(2\partial^{\lambda}\nc{A}^{\nu}-\partial^{\nu}\nc{A}^{\lambda})-\nc{A}^{\lambda}\left(2\partial^{\alpha}\nc{A}^{\nu}-\partial^{\nu}\nc{A}^{\alpha}\right)\right] = 0.
\end{split}
\end{equation}

Next we compute $\partial_{\mu}\nc{M}^{\mu\lambda\rho}$ from the definition \eqref{9181-lo}. Using the equation of motion \eqref{eom1-lo} and
\begin{equation}\label{lo-101}
\frac{\partial \nc{\mathscr{L}}}{\partial(\partial_{\mu}\nc{A}_{\lambda})} = \partial^{\lambda}\nc{A}^{\mu}-\partial^{\mu}\nc{A}^{\lambda}-\theta^{\alpha\beta}\partial_{\alpha}\nc{A}^{\mu}\partial_{\beta}\nc{A}^{\lambda}-\theta^{\mu\beta}\partial_{\beta}\nc{A}^{\nu}\left(\partial^{\lambda}\nc{A}_{\nu}-\partial_{\nu}\nc{A}^{\lambda}\right),
\end{equation}
it follows from \eqref{9181-lo} that
\begin{equation}\label{find2nc}
\partial_{\mu}\nc{M}^{\mu\lambda\rho} = \left(\partial_{\mu}\nc{A}_{\nu}-\partial_{\nu}\nc{A}_{\mu}\right)\left({\theta^{\lambda}}_{\alpha}\partial^{\alpha}\nc{A}^{\mu}\partial^{\rho}\nc{A}^{\nu}-{\theta^{\rho}}_{\alpha}\partial^{\alpha}\nc{A}^{\mu}\partial^{\lambda}\nc{A}^{\nu}\right),
\end{equation}
which shows that the criterion \eqref{crit2nc} is satisfied and not \eqref{crit22nc}. Thus, the action \eqref{nl-105} is invariant under Lorentz transformations only when $\theta$ transforms as a tensor, which is like the case of noncommutative electrodynamics in usual variables, considered in the previous section.

We shall now establish a connection between the two descriptions of noncommutative electrodynamics considered here and in the previous section. The Lagrangian densities in the two formulations are related by the map
\begin{equation}\label{LLhat}
\nc{\mathscr{L}} = \mathscr{L}+\frac{1}{4}\theta^{\alpha\beta}\partial_{\beta}\left(A_{\alpha}F_{\mu\nu}F^{\mu\nu}\right).
\end{equation}
Since $\nc{\mathscr{L}}$ and $\mathscr{L}$ differ by a total-derivative term, we have $\nc{S} = S$.

Now we shall find the maps between $T^{\mu\nu}$ and $\nc{T}^{\mu\nu}$ as well as between $M^{\mu\lambda\rho}$ and $\nc{M}^{\mu\lambda\rho}$. First we apply the Seiberg--Witten map \eqref{101-lo} on the right-hand side of Eq.~\eqref{lo-101} and take into account Eq.~\eqref{1191} to get
\begin{equation}\label{lo-101a}
\frac{\partial \nc{\mathscr{L}}}{\partial(\partial_{\mu}\nc{A}_{\lambda})} = \frac{\partial\mathscr{L}}{\partial(\partial_{\mu}A_{\lambda})}+\theta^{\alpha\mu}F^{\lambda\sigma}\partial_{\sigma}A_{\alpha}+\theta^{\alpha\lambda}F^{\sigma\mu}F_{\sigma\alpha}-\theta^{\alpha\beta}\partial_{\beta}(A_{\alpha}F^{\lambda\mu})+\frac{1}{4}\theta^{\lambda\mu}F_{\kappa\sigma}F^{\kappa\sigma}.
\end{equation}
Using the maps \eqref{101-lo}, \eqref{LLhat} and \eqref{lo-101a}, we get a map\footnote{A similar map among the symmetric energy--momentum tensors is defined in Ref.~\cite{BLY}. Note that the energy--momentum tensors considered here follow from Noether's prescription.} between $\nc{T}^{\mu\nu}$ \eqref{914-lo} and  $T^{\mu\nu}$ \eqref{914'}:
\begin{equation}\label{TThat}
\begin{split}
\nc{T}^{\mu\nu}&= T^{\mu\nu}+\theta^{\alpha\mu}\left(F^{\lambda\sigma}\partial_{\sigma}A_{\alpha}\partial^{\nu}A_{\lambda}+\frac{1}{4}F_{\lambda\rho}F^{\lambda\rho}\partial^{\nu}A_{\alpha}\right)\\
&\quad{}+\theta^{\alpha\beta}\left[\frac{1}{2}F^{\lambda\mu}\partial_{\lambda}A_{\alpha}\partial^{\nu}A_{\beta}-\partial_{\beta}\left(A_{\alpha}F^{\lambda\mu}\right)\partial^{\nu}A_{\lambda}-\frac{1}{4}\eta^{\mu\nu}\partial_{\beta}\left(A_{\alpha}F_{\lambda\rho}F^{\lambda\rho}\right)\right.\\
&\qquad\qquad\left.{}-\frac{1}{2}A_{\alpha}F^{\lambda\mu}\partial^{\nu}\left(\partial_{\beta}A_{\lambda}+F_{\beta\lambda}\right)\right].
\end{split}
\end{equation}
Similarly, using the maps \eqref{101-lo}, \eqref{lo-101a} and \eqref{TThat}, we get a map between $\nc{M}^{\mu\lambda\rho}$ \eqref{9181-lo} and  $M^{\mu\lambda\rho}$ \eqref{9181'}:
\begin{equation}\label{MMhat}
\nc{M}^{\mu\lambda\rho} = M^{\mu\lambda\rho}+M_{(\theta)}^{\mu\lambda\rho}-M_{(\theta)}^{\mu\rho\lambda},
\end{equation}
where
\begin{equation}\label{Mth-mlr}
\begin{split}
M_{(\theta)}^{\mu\lambda\rho}&=\theta^{\lambda\alpha}F^{\mu\sigma}F_{\sigma\alpha}A^{\rho}+\frac{1}{4}\theta^{\lambda\mu}A^{\rho}F_{\kappa\sigma}F^{\kappa\sigma}\\
&\quad{}+\theta^{\alpha\mu}\left[F^{\lambda\sigma}A^{\rho}\partial_{\sigma}A_{\alpha}-x^{\rho}\left(F^{\kappa\sigma}\partial_{\sigma}A_{\alpha}\partial^{\lambda}A_{\kappa}+\frac{1}{4}F_{\kappa\sigma}F^{\kappa\sigma}\partial^{\lambda}A_{\alpha}\right)\right]\\&\quad{}-\theta^{\alpha\beta}\left[A^{\rho}\partial_{\beta}\left(A_{\alpha}F^{\lambda\mu}\right)+\frac{1}{2}F^{\lambda\mu}A_{\alpha}\left(\partial_{\beta}A^{\rho}+{F_{\beta}}^{\rho}\right)\right.\\&\qquad\qquad\left.{}+x^{\rho}\left(\frac{1}{2}F^{\sigma\mu}\partial_{\sigma}A_{\alpha}\partial^{\lambda}A_{\beta}-\partial^{\lambda}A_{\sigma}\partial_{\beta}\left(A_{\alpha}F^{\sigma\mu}\right)-\frac{1}{4}\eta^{\mu\lambda}\partial_{\beta}\left(A_{\alpha}F_{\kappa\sigma}F^{\kappa\sigma}\right)\right.\right.\\&\qquad\qquad\qquad\quad\left.\left.{}-\frac{1}{2}A_{\alpha}F^{\sigma\mu}\partial^{\lambda}\left(\partial_{\beta}A_{\sigma}+F_{\beta\sigma}\right)\right)\right].
\end{split}
\end{equation}

It follows from Eq.~\eqref{TThat} that
\begin{equation}\label{lo-102}
\partial_{\mu}\nc{T}^{\mu\nu} = \partial_{\mu}T^{\mu\nu},
\end{equation}
where we have used the equation of motion, $\partial_{\mu}F^{\mu\nu}+O(\theta) = 0$. This shows the compatibility of the criteria for translational invariance in the two descriptions, Eqs.~\eqref{916-2}, \eqref{crit1n} and \eqref{916-lo}.

Next we show the compatibility of the criteria for Lorentz invariance. It follows from Eq.~\eqref{MMhat} that
\begin{equation}\label{lo-103}
\begin{split}
\partial_{\mu}\nc{M}^{\mu\lambda\rho} &= \partial_{\mu}M^{\mu\lambda\rho}+\theta^{\alpha\lambda}\left(F^{\sigma\mu}F_{\sigma\alpha}\partial_{\mu}A^{\rho}-F^{\sigma\mu}\partial_{\sigma}A_{\alpha}\partial^{\rho}A_{\mu}+\frac{1}{4}F_{\kappa\sigma}F^{\kappa\sigma}{F^{\rho}}_{\alpha}\right)\\&\quad{}-\theta^{\alpha\rho}\left(F^{\sigma\mu}F_{\sigma\alpha}\partial_{\mu}A^{\lambda}-F^{\sigma\mu}\partial_{\sigma}A_{\alpha}\partial^{\lambda}A_{\mu}+\frac{1}{4}F_{\kappa\sigma}F^{\kappa\sigma}{F^{\lambda}}_{\alpha}\right),
\end{split}
\end{equation}
where again the equation of motion, $\partial_{\mu}F^{\mu\nu}+O(\theta) = 0$, has been used. Now we use the maps \eqref{101-lo} and \eqref{lo-103} on the left-hand side of Eq.~\eqref{crit2nc} to obtain
\begin{equation}\label{lo-104}
\begin{split}
&\partial_{\mu}\nc{M}^{\mu\lambda\rho}-\left(\partial_{\mu}\nc{A}_{\nu}-\partial_{\nu}\nc{A}_{\mu}\right)\left({\theta^{\lambda}}_{\alpha}\partial^{\alpha}\nc{A}^{\mu}\partial^{\rho}\nc{A}^{\nu}-{\theta^{\rho}}_{\alpha}\partial^{\alpha}\nc{A}^{\mu}\partial^{\lambda}\nc{A}^{\nu}\right)\\ &= \partial_{\mu}M^{\mu\lambda\rho} - {\theta^{\lambda}}_{\alpha}F_{\mu\nu}\left(F^{\mu\alpha}F^{\nu\rho}+\frac{1}{4}F^{\mu\nu}F^{\rho\alpha}\right)+{\theta^{\rho}}_{\alpha}F_{\mu\nu}\left(F^{\mu\alpha}F^{\nu\lambda}+\frac{1}{4}F^{\mu\nu}F^{\lambda\alpha}\right).
\end{split}
\end{equation}
Thus, the left-hand side of criterion \eqref{crit2nc} goes over to the left-hand side of criterion \eqref{407n} under the Seiberg--Witten maps, which shows the compatibility of the two criteria for Lorentz invariance when $\theta$ transforms as a tensor. Turning to the case when $\theta$ does not transform, we now apply the maps \eqref{101-lo} and \eqref{lo-103} on the left-hand side of the criterion \eqref{crit22nc}:
\begin{equation}\label{lo-105}
\begin{split}
&\partial_{\mu}\nc{M}^{\mu\lambda\rho}-\frac{1}{2}{\theta^{\lambda}}_{\alpha}\left(\partial_{\mu}\nc{A}_{\nu}-\partial_{\nu}\nc{A}_{\mu}\right)\partial^{\mu}\left[\nc{A}^{\alpha}\left(2\partial^{\rho}\nc{A}^{\nu}-\partial^{\nu}\nc{A}^{\rho}\right)-\nc{A}^{\rho}\left(2\partial^{\alpha}\nc{A}^{\nu}-\partial^{\nu}\nc{A}^{\alpha}\right)\right]\\
&\quad{}+\frac{1}{2}{\theta^{\rho}}_{\alpha}\left(\partial_{\mu}\nc{A}_{\nu}-\partial_{\nu}\nc{A}_{\mu}\right)\partial^{\mu}\left[\nc{A}^{\alpha}\left(2\partial^{\lambda}\nc{A}^{\nu}-\partial^{\nu}\nc{A}^{\lambda}\right)-\nc{A}^{\lambda}\left(2\partial^{\alpha}\nc{A}^{\nu}-\partial^{\nu}\nc{A}^{\alpha}\right)\right]\\
&= \partial_{\mu}M^{\mu\lambda\rho}+\frac{1}{4}{\theta^{\lambda}}_{\alpha}\left[\partial^{\alpha}\left(A^{\rho}F_{\kappa\sigma}F^{\kappa\sigma}\right)-\partial^{\rho}\left(A^{\alpha}F_{\kappa\sigma}F^{\kappa\sigma}\right)\right]\\
&\quad{}-\frac{1}{4}{\theta^{\rho}}_{\alpha}\left[\partial^{\alpha}\left(A^{\lambda}F_{\kappa\sigma}F^{\kappa\sigma}\right)-\partial^{\lambda}\left(A^{\alpha}F_{\kappa\sigma}F^{\kappa\sigma}\right)\right].
\end{split}
\end{equation}
Thus, the left-hand side of criterion \eqref{crit22nc} goes over to the left-hand side of criterion \eqref{crit2n} up to total-derivative terms. The origin of these total-derivative terms is presumably due to the fact that $\nc{\mathscr{L}}$ and $\mathscr{L}$ are not exactly equal but differ by a total-derivative term, Eq.\eqref{LLhat}.

We shall now show that using the Noether charges
\begin{equation}\label{gen-lo}
\nc{P}^{\mu} = \int\!\mathrm{d}^{3}x\,\nc{T}^{0\mu}, \quad \nc{J}^{\mu\nu} = \int\!\mathrm{d}^{3}x\,\nc{M}^{0\mu\nu},
\end{equation}
and the canonical equal-time Poisson brackets $\{\nc{A}_{\mu}(t, \mathbf{x}), \nc{\pi}^{\nu}(t, \mathbf{y})\} = \delta_{\mu}^{\nu}\delta^{3}(\mathbf{x}-\mathbf{y})$, we can generate the transformations of the dynamical variables $\nc{A}_i$ and $\nc{\pi}_i$:
\begin{equation}\label{AQL-lo}
\left\{\nc{A}_{i}, \nc{Q}_{V}\right\} = \mathcal{L}_{V}\nc{A}_{i}, \quad
\left\{\nc{\pi}^{i}, \nc{Q}_{V}\right\}= \mathcal{L}_{V}\nc{\pi}^{i}.
\end{equation}

The canonical momenta of the theory are
\begin{gather}
\label{pi02-lo}
\nc{\pi}^{0} = -\theta^{0i}\partial_{i}\nc{A}_{j}\left(\partial^{0}\nc{A}^{j}-\partial^{j}\nc{A}^{0}\right),\\
\label{pii2-lo}
\begin{split}
\nc{\pi}^{i} &=\partial^{i}\nc{A}^{0}-\partial^{0}\nc{A}^{i}-\theta^{kl}\partial_{k}\nc{A}^{0}\partial_{l}\nc{A}^{i}\\
&\quad{}-\theta^{0l}\left(\partial_{0}\nc{A}^{0}\partial_{l}\nc{A}^{i}-2\partial_{l}\nc{A}^{0}\partial_{0}\nc{A}^{i}+\partial_{l}\nc{A}_{0}\partial^{i}\nc{A}^{0}+\partial^{i}\nc{A}^{k}\partial_{l}\nc{A}_{k}-\partial^{k}\nc{A}^{i}\partial_{l}\nc{A}_{k}\right).
\end{split}
\end{gather}
As in the previous section, here also we set $\theta^{0i}=0$, so that the above definitions simplify to
\begin{gather}
\label{pi0-lo}
\nc{\pi}^{0} = 0,\\
\label{pii-lo}
\nc{\pi}^{i} =\partial^{i}\nc{A}^{0}-\partial^{0}\nc{A}^{i}-\theta^{kl}\partial_{k}\nc{A}^{0}\partial_{l}\nc{A}^{i}.
\end{gather}
It follows from the definitions \eqref{914-lo}, \eqref{9181-lo} and \eqref{pii-lo} that
\begin{gather}
\label{T00-lo}
\begin{split}
\nc{T}^{00} &= \nc{\pi}^{i}\partial_{i}\nc{A}^{0}-\frac{1}{2}\nc{\pi}^{i}\nc{\pi}_{i}-\frac{1}{2}\partial_{i}\nc{A}_{j}\left(\partial^{i}\nc{A}^{j}-\partial^{j}\nc{A}^{i}\right)\\
&\quad{}-\theta^{kl}\left[\nc{\pi}_{i}\partial_{k}\nc{A}^{0}\partial_{l}\nc{A}^{i}+\partial_{k}\nc{A}^{i}\partial_{l}\nc{A}^{j}\partial_{i}\nc{A}_{j}\right],
\end{split}\\
\label{T0i-lo}
\nc{T}^{0i} = \nc{\pi}^{j}\partial^{i}\nc{A}_{j},\\
\label{M00i-lo}
\nc{M}^{00i} = -\nc{T}^{00}x^{i}-\nc{\pi}^{i}\nc{A}^{0}+x^{0}\nc{\pi}^{j}\partial^{i}\nc{A}_{j},\\
\label{M0ij-lo}
\nc{M}^{0ij} = \nc{\pi}^{i}\nc{A}^{j}-x^{j}\nc{\pi}^{k}\partial^{i}\nc{A}_{k}-\nc{\pi}^{j}\nc{A}^{i}+x^{i}\nc{\pi}^{k}\partial^{j}\nc{A}_{k}.
\end{gather}
After some algebra, we find that
\begin{equation}
\label{APL-lo}
\left\{\nc{A}_{i}, \nc{P}_{\mu}\right\} = \mathcal{L}_{\partial_{\mu}}\nc{A}_{i},\quad
\left\{\nc{A}_{i}, \nc{J}_{\mu\nu}\right\} = \mathcal{L}_{x_{[\mu}\partial_{\nu]}}\nc{A}_{i},
\end{equation}
and likewise for $\nc{\pi}_{i}$, which proves Eq.~\eqref{AQL-lo}. We also find that
\begin{gather}
\label{PP-lo}
\left\{\nc{P}_{i}, \nc{P}_{j}\right\} = 0,\\
\label{PJ-lo}
\left\{\nc{P}_{i}, \nc{J}_{kl}\right\} = \eta_{ik}\nc{P}_{l}-\eta_{il}\nc{P}_{k},\\
\label{JJ-lo}
\left\{\nc{J}_{ij}, \nc{J}_{kl}\right\} = \eta_{jk}\nc{J}_{il}+\eta_{il}\nc{J}_{jk}-\eta_{ik}\nc{J}_{jl}-\eta_{jl}\nc{J}_{ik},
\end{gather}
from where it follows that
\begin{equation}\label{QQ-lo}
\left\{\nc{Q}_{U}, \nc{Q}_{V}\right\} = \nc{Q}_{[U,V]},
\end{equation}
where we have restricted to kinematical generators ($\nc{P}_{i}$ and $\nc{J}_{ij}$) only.
Thus we see that although $\partial_{0}\nc{Q}_{V} \neq 0$ (for rotations and boosts), we still have Eqs.~\eqref{AQL-lo} and \eqref{QQ-lo}. This is necessary for establishing the dynamical consistency of the transformations.

Finally, we would like to mention that for the choice \eqref{lo-113} of $\theta$, Eq.~\eqref{find2nc} gives $\partial_{\mu}\nc{M}^{\mu 23}=0$ and $\partial_{\mu}\nc{M}^{\mu 01}=0$. The criterion \eqref{crit22nc} for Lorentz invariance when $\theta$ does not transform is not compatible with Eq.~\eqref{find2nc} in general. However, for this particular choice of $\theta$ the criterion \eqref{crit22nc} also gives $\partial_{\mu}\nc{M}^{\mu 23}=0$ and $\partial_{\mu}\nc{M}^{\mu 01}=0$. Thus, Lorentz invariance is partially restored.


\section{\label{sec:conclu}Conclusions}

We have derived, starting from a first-principle Noether-like approach, criteria for preserving Poincar\'e invariance in a noncommutative gauge theory with a constant noncommutativity parameter $\theta$. The criterion for translational invariance was the same irrespective of whether $\theta$ transformed as a second-rank tensor or was the same constant in all frames. This criterion was then shown to hold by performing an explicit check. Thus, as expected, translational invariance was valid.

The issue of Lorentz invariance (invariance under rotations and boosts) was quite subtle. Here we found distinct criteria depending on the nature of transformation of $\theta$. An explicit check using the equations of motion confirmed the particular criterion for Lorentz invariance when $\theta_{\mu\nu}$ transformed as covariant second-rank tensor. Thus Lorentz invariance was preserved only for a transforming $\theta$.

We have also shown that all the transformations are dynamically consistent. The Noether charges generated the appropriate transformations on the phase-space variables. These charges also satisfied the desired Lie brackets among themselves.

The complete analysis was done in both the commutative and noncommutative descriptions. By the use of suitable Seiberg--Witten-type maps, compatibility among the results found in the two descriptions was established.

The criteria for Lorentz invriance found here were also consistent with the fact that, for a constant non-transforming $\theta$ having special values, the symmetry group breaks down to $\mathrm{SO}(1,1)\times\mathrm{SO}(2)$, a subgroup of the Lorentz group.

The present analysis fits in with the general notions of observer versus particle Lorentz transformations. As is known, usually (without a background) these two approaches to Lorentz symmetry agree. In the presence of a background, this equivalence fails since the background (here $\theta_{\mu\nu}$) transforms as a tensor under observer Lorentz transformations but as a set of scalars under particle Lorentz transformations. The effect of observer and particle Lorentz transformations was captured here by the distinct set of criteria (Eqs.~\eqref{407n} and \eqref{crit2n} in the commutative description and Eqs.~\eqref{crit2nc} and \eqref{crit22nc} in the noncommutative description) obtained for a transforming or a non-transforming $\theta$. Lorentz symmetry was preserved only for a transforming $\theta$ which conforms to observer Lorentz transformations.

The analysis of Lorentz symmetry in the presence of the background field $\theta$ seems to parallel the discussion of gauge symmetry\footnote{For a detailed study of the connection between Lorentz and gauge symmetries in the Maxwell theory, see Ref.~\cite{Wein}.} in the presence of a background magnetic field $B$.\footnote{Indeed $\theta$ can be regarded as the inverse of $B$.} In the present treatment, Lorentz symmetry of the action is preserved although there may not be a conserved generator.\footnote{Note however that the generators are dynamically consistent as shown, for instance, in Eqs.~\eqref{AQLn}, \eqref{QQn}, \eqref{AQL-lo} and \eqref{QQ-lo}.} Likewise, gauge symmetry of the action, say for a particle moving in the presence of background magnetic field, is preserved although a generator, like the Gauss operator, does not exist, since there is no dynamical piece for the gauge field.

Finally, we mention that the present analysis refers to the standard realization of Poincar\'e symmetry over trivial co-commutative Hopf algebra of fields. The principal result, that the Lorentz symmetry is restored for a suitably transforming $\theta$, is expected within such a scheme. Recently it has been shown \cite{CKNT, CPT} that for constant $\theta$, an explicit twisted Poincar\'e symmetry is realized within the twisted Hopf algebra of fields.


\section*{Acknowledgments}
K. K. thanks the Council of Scientific and Industrial Research (CSIR), Government of India for financial support.



\end{document}